\documentclass[preprint,12pt]{elsarticle}

\usepackage{epsfig}

\journal{Future Generation Computer Systems}

\begin{document}

\begin{frontmatter}

\title{Cloudpress 2.0: A MapReduce Approach for News Retrieval on the Cloud}

\author{Arockia Anand Raj\corref{cor1}}
\ead{arockiaanandraj@hotmail.com}
\address{Department of Information Science and Technology, Anna University, Chennai - 600025 India}

\author{T. Mala\corref{cor1}}
\ead{malanehru@annauniv.edu}
\address{Department of Information Science and Technology, Anna University, Chennai - 600025 India}
\cortext[cor1]{Corresponding author}

\begin{abstract}
In this era of the Internet, the amount of news articles added every minute of everyday is humongous. As a result of this explosive amount of news articles, news retrieval systems are required to process the news articles frequently and intensively. The news retrieval systems that are in-use today are not capable of coping up with these data-intensive computations. \textit{Cloudpress 2.0} presented here, is designed and implemented to be scalable, robust and fault tolerant. It is designed in such a way that, all the processes involved in news retrieval such as fetching, pre-processing, indexing, storing and summarizing, exploit MapReduce paradigm and use the power of the Cloud computing. It uses novel approaches for parallel processing, for storing the news articles in a distributed database and for visualizing them as a 3D visual. It uses Lucene-based indexing for efficient and faster retrieval. It also includes a novel query expansion feature for searching the news articles. \textit{Cloudpress 2.0} also allows on-the-fly, extractive summarization of news articles based on the input query.
\end{abstract}


\end{frontmatter}

\section{Introduction}
\label{intro}

The main functions of a news retrieval system involve fetching, processing and retrieval of news articles in any format such as text or image or video or audio or combination of any of those. News retrieval systems which are used nowadays, does not fully exploit parallelism to increase performance and to decrease time taken for retrieval. Furthermore, they are not designed to be powered by Cloud technology and to follow MapReduce approach.

Generally, news articles have less number of mistakes as they are written by professionals and they are often pieces of larger stories which contain rich information about the people and places involved. However, they need to be fetched from the Internet, pre-processed, indexed and stored into the database. Often, due to the fact that amount of news articles is evergrowing, a distributed database is preferred for storage. MapReduce framework can be used to split each and every task into sub-tasks and then assign them to different worker nodes present in the Cloud for parallel processing. This can reduce the processing time, greatly.

The paper encompasses a brief literature survey in the field of news retrieval systems, parallel crawlers, news visualization and MapReduce programming model in Section 2. In Section 3, an overall architecture of the news retrieval system presented in this paper, is introduced. In Section 4, the implementation details of all the sub components comprising the system is explained. In Section 5, performance evaluation of the system is dealt with and finally, in Section 6, the concluding remarks with potential future enhancements are presented.

\section{Literature Survey}

\subsection{News Retrieval System}

A novel design of a \textit{news retrieval tool} is introduced in \cite{nrt}. It makes use of an existing database of some newspapers such as Times. Two algorithms are presented in \cite{nrt}, namely, conflation algorithm and relevance feedback algorithm. The conflation algorithm strips the suffixes off words, leaving a root stem. The rules used in this algorithm are very simple but the tool performs as well as many of the more complex algorithms that have been developed. The relevance feedback algorithm enables the user to add news articles which are considered by the user to be relevant, to the input query and the system analyses it to extract keywords from it. This algorithm releases the burden of having to think up new words for the input query.

A \textit{news filtering and summarization system} presented in \cite{nfas} can automatically recognize Web news pages, retrieve each news page's title and news content and extract key phrases. The key phrases extraction described in \cite{nfas} performs better than the methods based on term frequency and lexical chains.

\subsection{Parallel Crawler}

A scalable, extensible Web crawler named \textit{Mercator} written entirely in Java, is discussed in \cite{merc}. It enumerates the major components of any scalable Web crawler, comments on alternatives and tradeoffs in their design, and describes the particular components used in Mercator. It also describes Mercator's support for extensibility and customizability.

A scalable, distributed crawler named \textit{UbiCrawler} introduced in \cite{ubi}, is characterized by platform independence, linear scalability, graceful degradation if any faults occur and it has an assignment function which partitions the domain to crawl. The limitations of handling large sets of data have been explained and the techniques to overcome them are also presented in \cite{ubi}.

\subsection{News Visualization}

A tree map visualization is used for visualizing news articles in \cite{newsmap} and it deals with some interactivity and abstraction issues as well. It focuses on the automatic generation of a hierarchical knowledge map called the NewsMap, based on online Chinese news, particularly the finance and health sections. The hierarchical knowledge map may be used as a tool for browsing business intelligence and medical knowledge hidden in news articles. \textit{NewsMap} employs an improved interface combining a 1D alphabetical hierarchical list and a 2D Self-Organizing Map (SOM) island display.

\subsection{MapReduce Paradigm}

The \textit{MapReduce} paradigm is explained in \cite{mapreduce} as a programming model and an associated implementation for processing and generating large datasets that can be applied to various real-world tasks. It is explained that users can specify the computation in terms of a map and a reduce function, and the underlying runtime system automatically parallelizes the computation across large-scale clusters of machines, handles machine failures and schedules inter-machine communication to make efficient use of the network and disks.

\section{Overall Architecture}

\begin{figure}[!h]
\centering
\epsfig{file=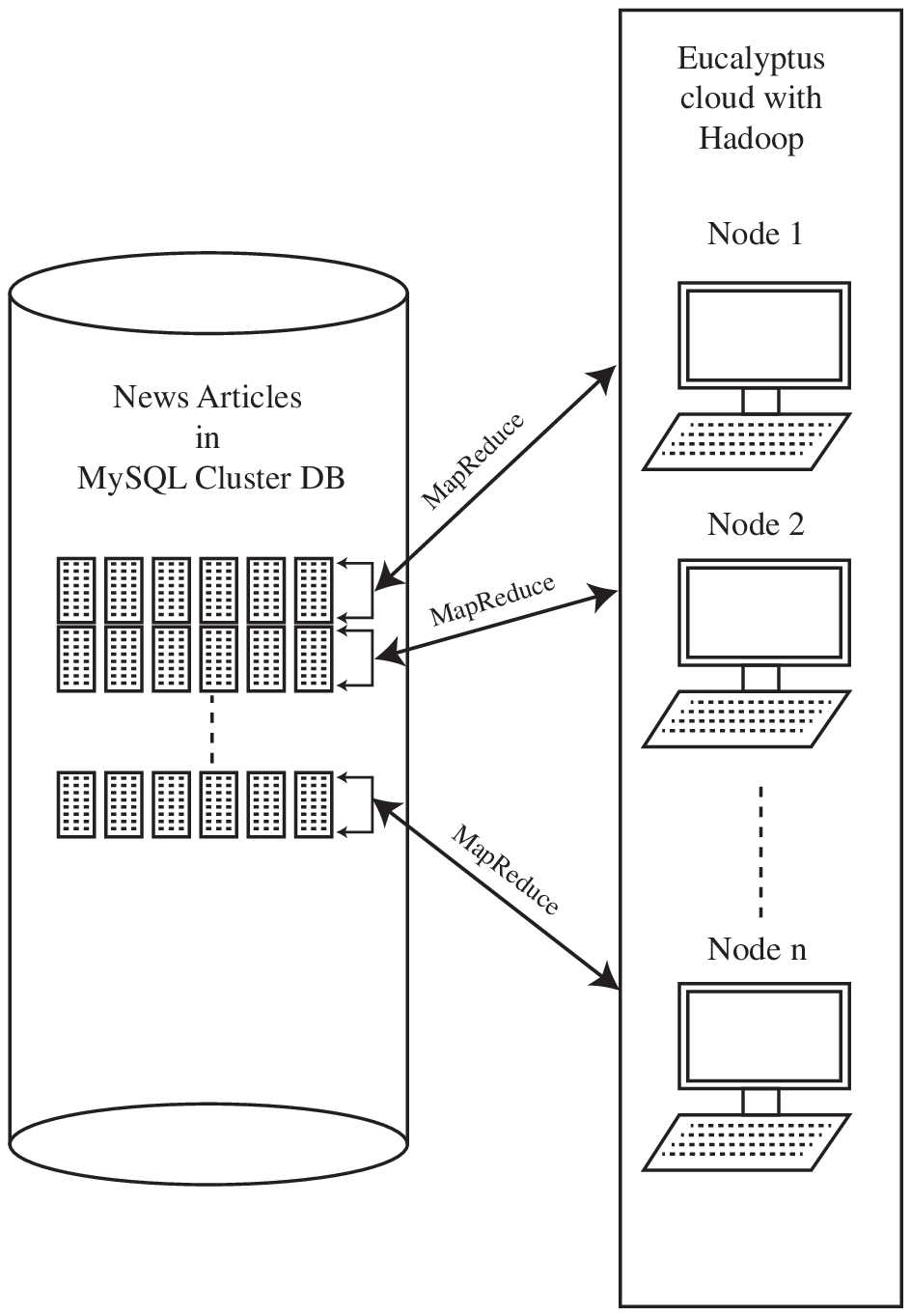, height=3.5in}
\caption{MapReduce processing approach}
\label{fig_mr}
\end{figure}

Every data-intensive process which takes place in the news retrieval system presented here, takes a MapReduce approach as shown in Figure~\ref{fig_mr}.

\begin{figure}[!ht]
\centering
\epsfig{file=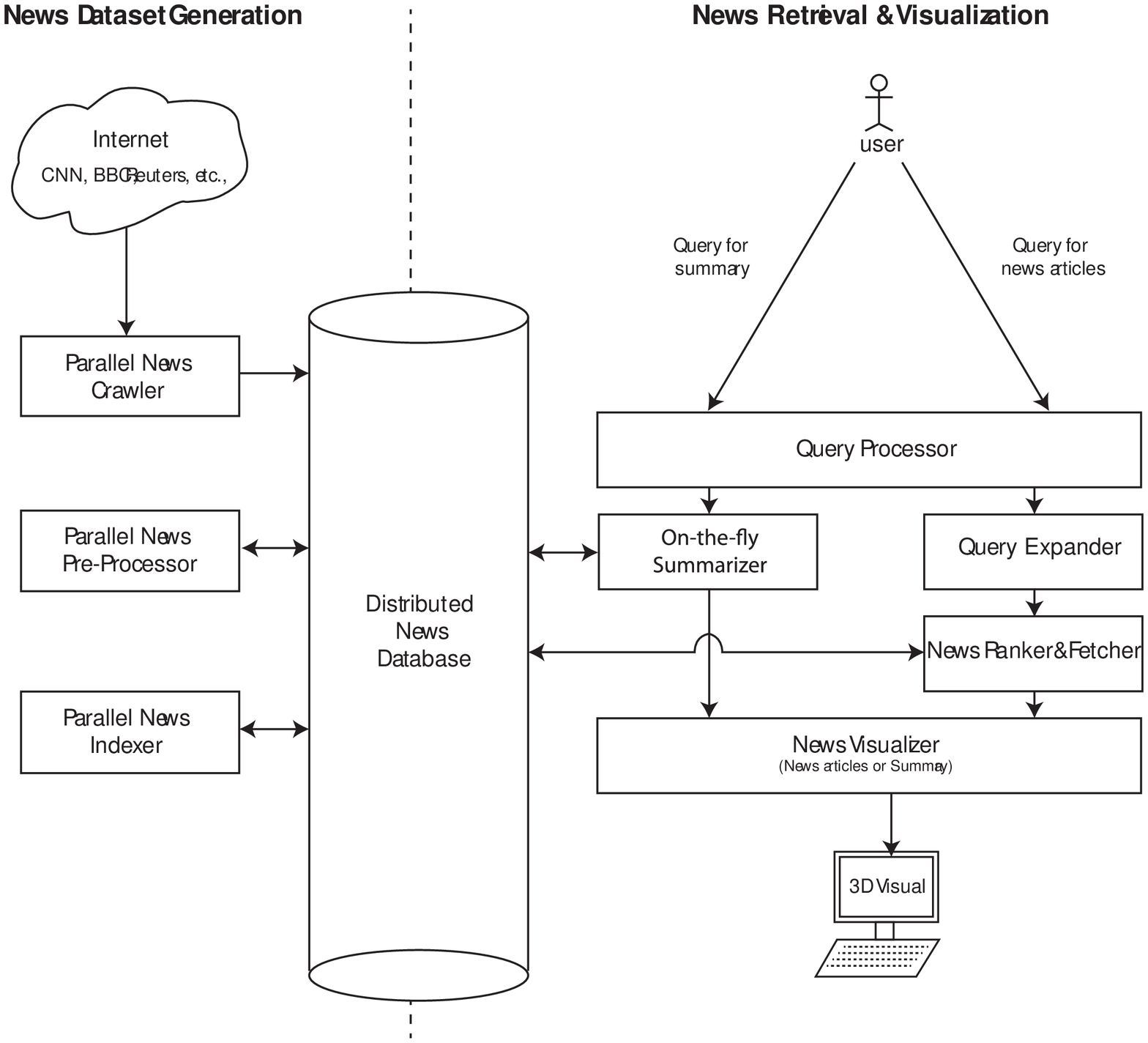, height=4in}
\caption{Architecture of Cloudpress 2.0}
\label{fig_arch}
\end{figure}

The overall architecture shown in Figure~\ref{fig_arch}, shows two functions of the news retrieval system. They are \textit{News Dataset Generation} and \textit{News Retrieval and Visualization}. \textit{News Dataset Generation} involves crawling, pre-processing, indexing and storing the news articles in a parallel fashion. \textit{News Retrieval and Visualization} involves query processing, query expansion, ranking, summarizing and retrieving news articles from the distributed database and finally, visualizing them as 3D visuals.

\section{Implementation}

All the processes involved in the news retrieval system presented in this paper are basically designed and implemented as maps and reduces which works on the \textit{Cloud using Hadoop framework}.

\subsection{Parallel News Crawler}

\begin{figure}[!h]
\centering
\epsfig{file=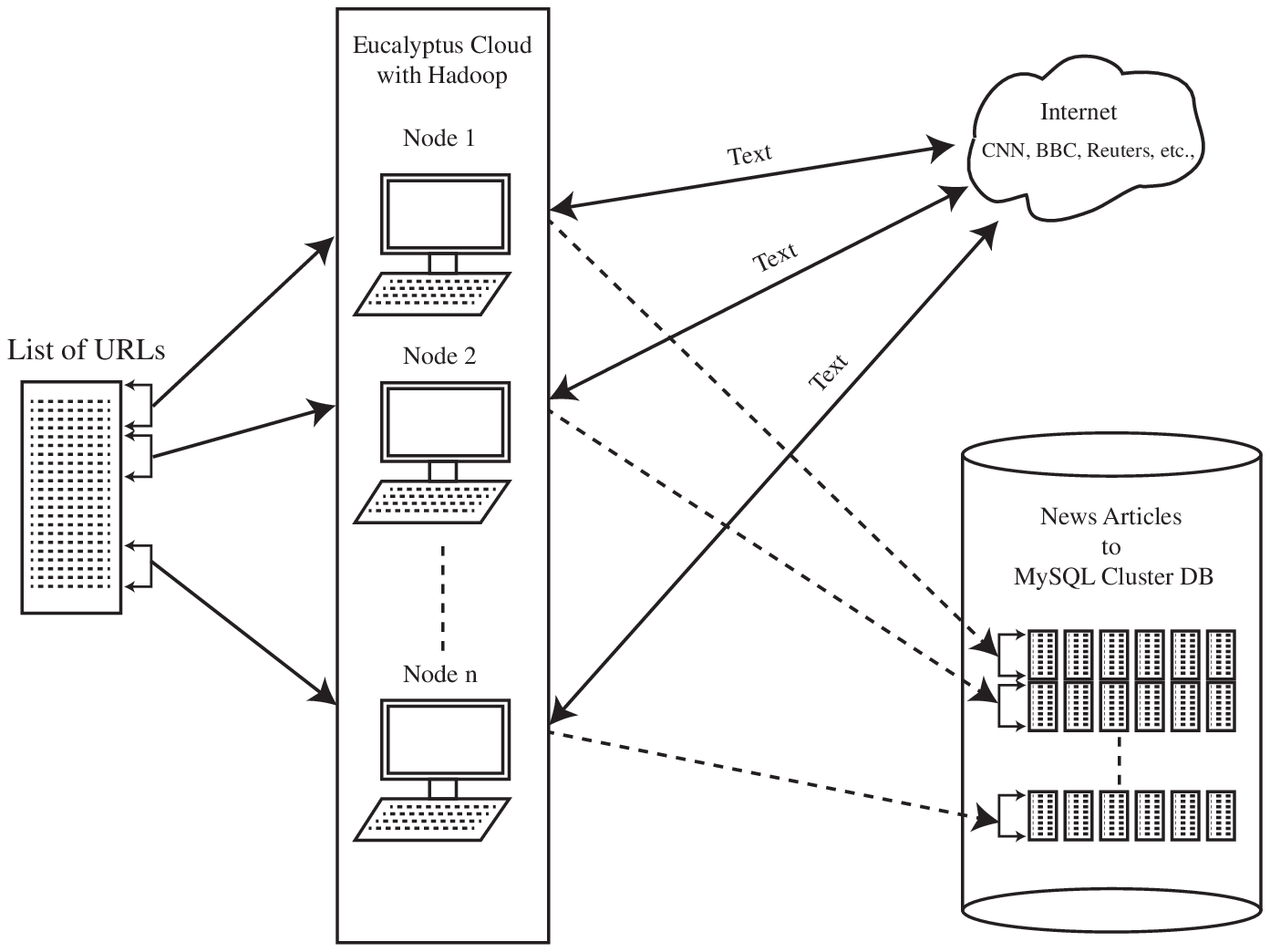, width=3.2in}
\caption{Working of the Parallel News Crawler}
\label{fig_crawl_arch}
\end{figure}

News articles are fetched from the Internet simultaneously by various worker nodes in the Cloud, by using this parallel news crawler. The parallel news crawler is implemented using \textit{JAVA} and \textit{Hadoop framework}. The list of input URLs which resides in \textit{Hadoop Distributed File System}(HDFS) is split among the worker nodes present in Cloud and each node crawls independently and stores the news articles into the distributed database. This is shown in the Figure~\ref{fig_crawl_arch}. If crawling a particular URL takes a longtime then instead of waiting for that node to finish crawling, a new Hadoop worker node can be instantiated and the next URL can be assigned to it, so as to keep the speed of the crawling high, compared to the other crawlers which are in-use today.

\begin{figure}[!h]
\centering
\epsfig{file=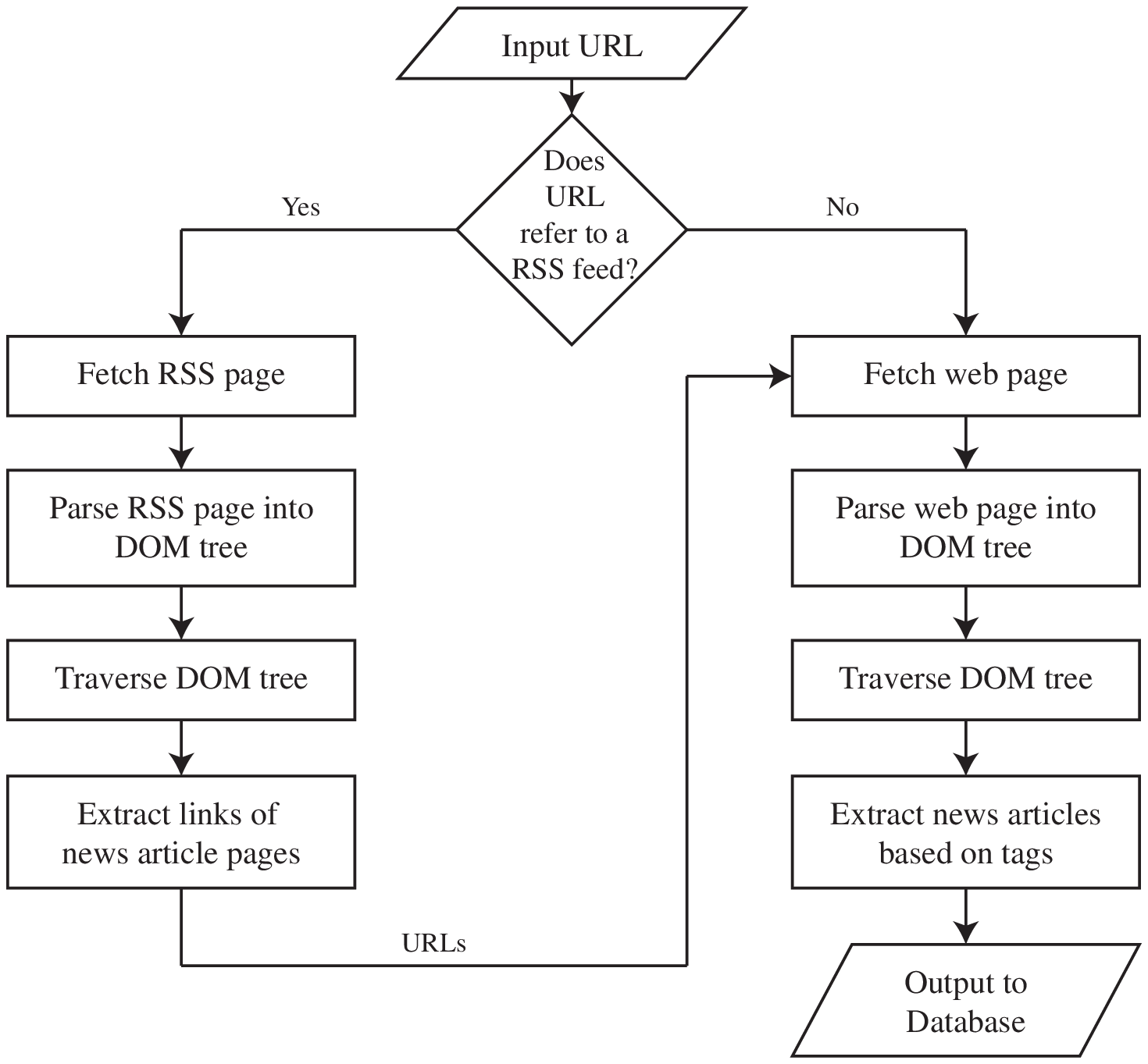, height=3in}
\caption{Steps involved in parallel news crawling in each node}
\label{fig_crawl}
\end{figure}

Basically, the nodes can be instantiated or terminated based on the workload and thereby exploiting the full power of the Cloud technology. Further, the steps involved in the process of crawling done by each worker node, are shown in Figure~\ref{fig_crawl}.

\subsection{Parallel News Pre-Processor}

Pre-processing starts with getting the input from the distributed database. Then the first and foremost process to be carried out is the \textit{tokenization} of the whole news article. In this step, all of the punctuation marks and symbols are removed by scanning each and every character in the news article. After that, the words (sets of characters separated by a space in-between) are then counted one-by-one to get the total \textit{word count}. Next, the \textit{stop words} like prepositions, articles and some commonly occurring words are removed by comparing the tokenized words of the news article with a standard stop word list. This stop word list is kept centrally in HDFS, since it is needed by all the nodes. At last, the words in the news article  are \textit{stemmed} using WordNet, which converts them to their root form without any prefix or suffix characters. The words which are not matching or not found in the WordNet like names of persons, places and numbers, are simply left to be the same. The completion of the stemming process also completes the full pre-processing stage. The pre-processed list of words are finally stored in the distributed database as \textit{Comma Separated Values}(CSVs).

\subsection{Parallel News Indexer}

The parallel news indexer is implemented in JAVA which creates two Lucene-based indexes. The pre-processed CSVs from the distributed database is fetched and supplied as an input to generate one index which is used for faster retrieval of news articles, if the input query is for exact matching query terms. Another index is created with the original news articles (as they were before preprocessing). This index is created with term vectors and positional offsets and is used for proximity search queries, that is, fetching news articles with two or more keywords occurring within a particular distance from each other. For example, for the query \textit{``war iraq"{\raise.17ex\hbox{$\scriptstyle\sim$}}2}, the resultant news articles would have both ``\textit{war}" and ``\textit{iraq}" within two words of each other.  This indexer also calculates and stores the \textit{Inverse Document Frequencies}($IDF$) of all the terms in the index. The $IDF$ of each word $w_i$ is given by (\ref{eqn_idf}),
\begin{equation}
\label{eqn_idf}
IDF_i = \log \frac {N}{n_i}
\end{equation}

where, $N$ = total number of articles in news database and  $n_i$ = number of news articles that contain the word $w_i$.

\subsection{Query Processor, Expander, Ranker and Fetcher}
\begin{figure}[!ht]
\centering
\epsfig{file=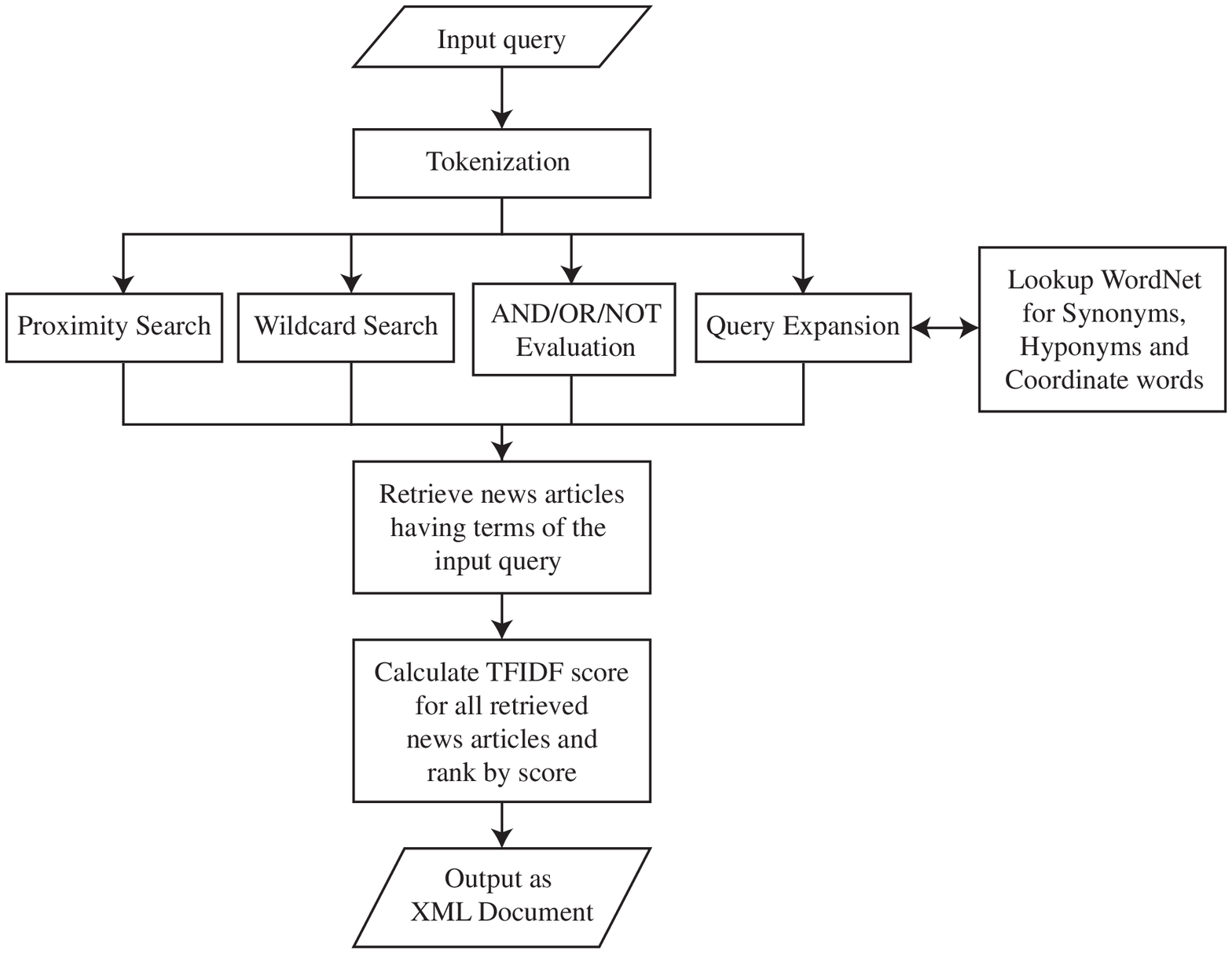, height=4in}
\caption{Steps involved in query processing and expanding in each node}
\label{fig_queryprocess}
\end{figure}

The basic steps involved in the query processing, expanding, ranking and fetching of news articles is shown in Figure~\ref{fig_queryprocess}. The input query can be for a \textit{summary} or \textit{all relevant news articles}. If it is for a \textit{summary}, automatically the query expansion feature is disabled, so as to obtain a relavant summary rather than a far-fetched one.
First, the input query from the user is \textit{tokenized} and the punctuations are removed, if necessary. The user can choose the way the input query needs to be processed. The user can choose whether \textit{AND/OR/NOT evaluation} or \textit{wildcard character matching} or the \textit{exact query term matching} needs to be done for retrieval of the news articles.

If the user opts for \textit{AND/OR/NOT evaluation}, then the tokenized query terms are checked if it matches `AND' or `OR' or `NOT'. If they do not match, then the query term can be directly used for retrieval by exact match. But if it matches any one of term, then corresponding outputs are obtained as follows:

\begin{itemize}
\item If it matches `AND' then, document IDs of the words occurring before and after `AND' are retrieved from inverted index and they are compared to get their intersection, which will give the final output (or)
\item If it matches `OR' then, document IDs of the words occurring before and after `OR' are retrieved from inverted index and combined (or)
\item If it matches `NOT' then, document IDs of the word occurring after `NOT' are retrieved from inverted index and the final output will be all documents IDs except these.
\end{itemize}

If the user opts for \textit{query expansion}, then the tokenized query term is looked up for its \textit{synonym}, \textit{hyponym} and \textit{coordinate terms} in WordNet. Then, all the document IDs of news articles containing any and all those terms are retrieved from the database.

If the user opts for \textit{wildcard character matching}, then the untokenized query term can be directly used for retrieval. For example, the query term like ` hel? ' or ` go* ' can be directly used to retrieve the document IDs. After the document IDs are obtained, their Term Frequency-Inverse Document Frequency($TFIDF$) score is calculated. $TFIDF$ score is given by (\ref{eqn_tfidf}),

\begin{equation}
\label{eqn_tfidf}
TFIDF_i = tf_i  \times \log \frac {N}{n_i}
\end{equation}

where,  $tf_i$ = frequency of the word $w_i$ in the given news article, $N$ = total number of articles in news database and $n_i$ = number of news articles that contain the word $w_i$.

Based on this score, the news articles are ranked. Until this, the process remains same for both an input query for summary and an input query for all relevant news articles. 

But after the ranking is done, 
\begin{itemize}
\item if the input query was for \textit{all relevant news articles} then, based on the $TFIDF$ score, they are ordered and corresponding news articles are retrieved one by one and written to an XML file in the same order. An example output XML file is shown in the Figure~\ref{fig_xml}.

\begin{figure*}[!ht]
\centering
\epsfig{file=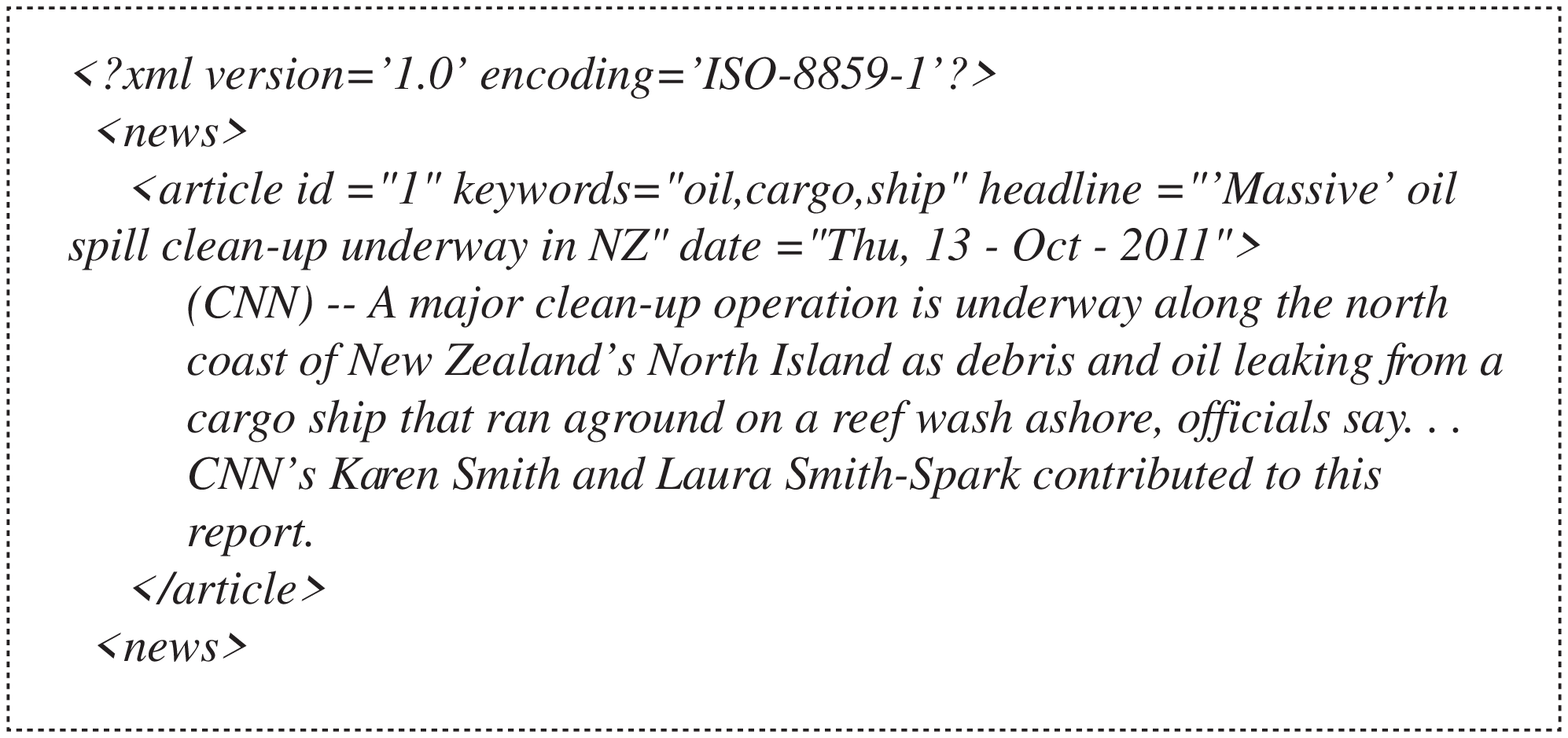, height=2in}
\caption{An example output XML file}
\label{fig_xml}
\end{figure*}

\item if the input query was for a \textit{summary} then the relevant news articles are fetched and then the steps shown in Figure~\ref{fig_summary} are followed to get a summary as an XML output file.

\begin{figure}[!ht]
\centering
\epsfig{file=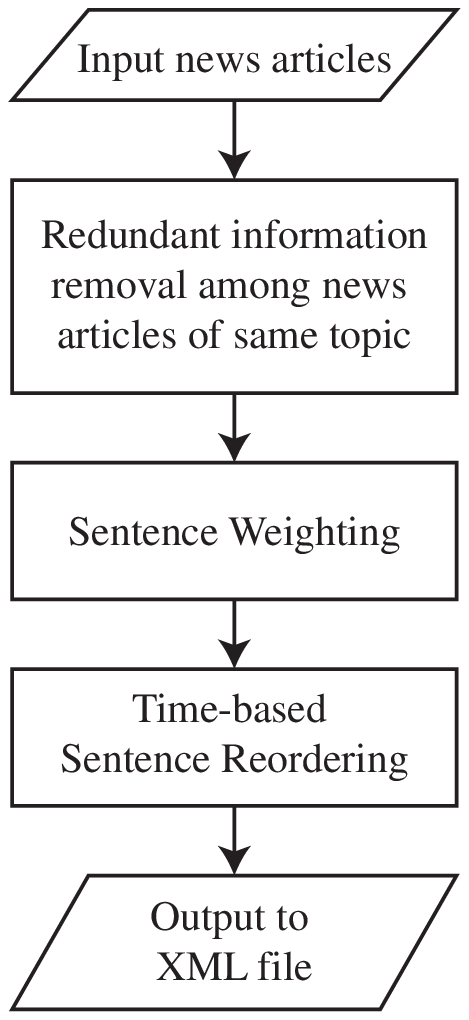, height=3.0in}
\caption{Steps involved in summary generation}
\label{fig_summary}
\end{figure}

\end{itemize}

\subsection{3D News Visualizer}
\begin{figure}[!h]
\centering
\epsfig{file=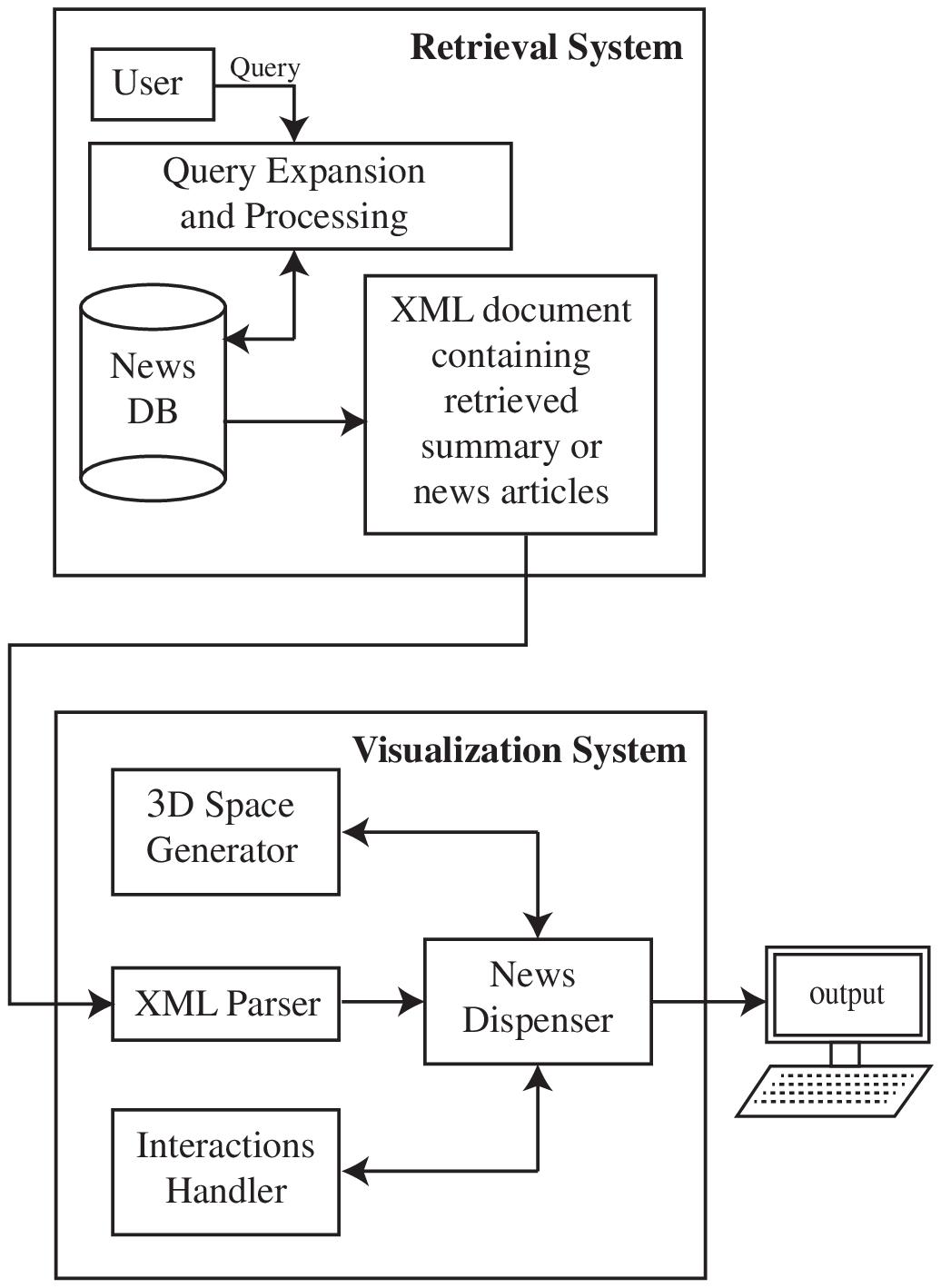, height=3.8in}
\caption{Architecture of retrieval and visualization modules}
\label{fig_visual}
\end{figure}

The architecture of the news retrieval and visualization modules is shown in the Figure~\ref{fig_visual}. In the \textit{3D News Visualizer}, the XML file created by the previous module is fed to the \textit{XML parser}, which then parses the tags contained in the XML document to get all the news articles iteratively and supplies it to the \textit{news dispenser}. Then, the \textit{3D space generator} creates a 3D screen space with a camera focusing on the front view. 

The \textit{news dispenser}, dynamically adds the news articles to various points in 3D space based on previously ranked order, in other words, the z-axis coordinate or depth value of each news article is determined by the order or the \textit{'id'} attribute in each \textit{'article'} tag in the XML. The \textit{'keywords'} attribute  in each of the \textit{'article'} tags is used to highlight the query terms previously given by the user as input to the retrieval system.

There are two views associated with each news article in 3D screen space. They are, namely, \textit{Title view} and \textit{Detailed view}. The \textit{default view} of the system is the \textit{Title view}. In \textit{Title view}, only the news article's ID, title and date of publishing will be visible. In \textit{Detailed view}, the news article's ID, title, date of publishing and the full news story in form of text with highlighted keywords is displayed.

The user-interface of 3D News Visualizer is implemented using \textit{Adobe Flash} and all the necessary event handling is done using \textit{ActionScript 3.0}. 3D News Visualizer has the following interactive features: \textit{Panning}, \textit{Zooming} and \textit{Selecting/Deselecting}.

\textit{Panning} can be performed by the use of arrow keys in the keyboard. In order to pan to the left direction, the user can press the left arrow key and likewise for right direction, the right arrow key can be used. Similarly, for top and bottom directions, the up and down arrow keys can be used. While in detailed view, the mouse pointer can be moved in the direction the user wishes, to pan and the corresponding panning will occur.

\textit{Zooming} is nothing but the increase or decrease of depth values and sizes of all the news articles, to simulate the effect of flying through 3D space. It is achieved by the use of scroll wheel in the mouse. Scrolling up achieves zoom-in and scrolling down achieves zoom-out. It can also be performed by double clicking the left button in the mouse, over a title far in 3D space.

\textit{Selecting/Deselecting} can be done by a left button click of the mouse, over the title of the news article. It toggles between title view and detailed view. Every time the user clicks, the corresponding news article's ID is displayed at left bottom of the screen. And at the right bottom of the screen, the total number of retrieved news articles is displayed.

\section{Performance Evaluation}

\subsection{Our Parallel Crawler vs Other Crawlers}

\begin{figure}[!h]
\centering
\epsfig{file=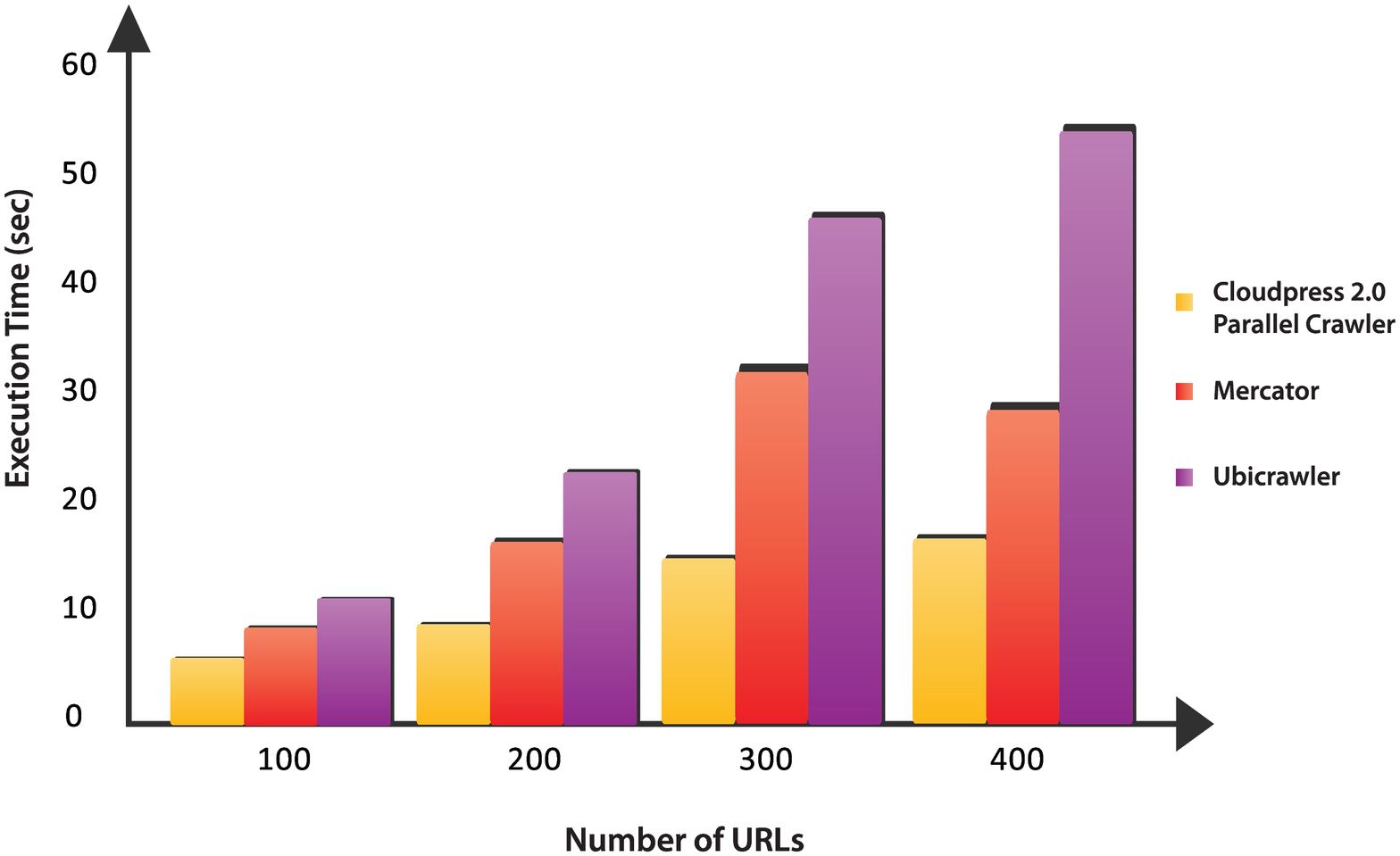, width=3.8in}
\caption{Execution Times of Our Parallel Crawler vs Other Crawlers}
\label{fig_crawlchart}
\end{figure}

In this evaluation, the parallel crawler's efficiency and performance is compared with other parallel crawlers such as Ubicrawler and Mercator. Our parallel crawler was made to execute in an eight node setup. For this evaluation, in each trail, the number of RSS feeds are increased. Each RSS feed contains about 100-120 links or URLs in them. The results are shown in the Figure~\ref{fig_crawlchart}.

The results show that, for any given number of RSS feeds, our parallel crawler is approximately twice as fast as the other crawlers. This behavior of the parallel crawler is due to its use of Map-Reduce approach coupled with the efficient usage of Cloud resources. The parallel crawler splits the input URLs evenly and then assigns them to each of the nodes to crawl simultaneously. If one URL takes more than a set time limit to be fully crawled, then a new Hadoop node is instantiated and the crawling of the next URL begins. This is followed until the rest of the URLs are crawled. But in the other crawlers, the crawling is limited by the number of nodes present at start of the crawling process and dynamic instantiation of nodes is not efficiently done. The crawler described in this paper has the ability to cater a fully variable workload with ease and yet consume only sufficient amount of computing resources  with full utilization.

\subsection{Normal Query vs Expanded Query}

For this evaluation, total number of news documents were increased at constant steps by adding news documents of random topics and each time the same query term is input to the system. At each trail, the query term is evaluated as a normal query and as an expanded query. 

For example, if the input query term is \textit{`kill'}, then, for once it is taken as a normal query and an exact match with the news articles in the distributed database is performed to retrieve the resultant news articles. Then the same query term \textit{`kill'} is expanded which gives additional terms like \textit{`out'} and \textit{`eliminate'}. This expanded query is then executed to retrieve the resultant news articles.

\begin{figure}[!h]
\centering
\epsfig{file=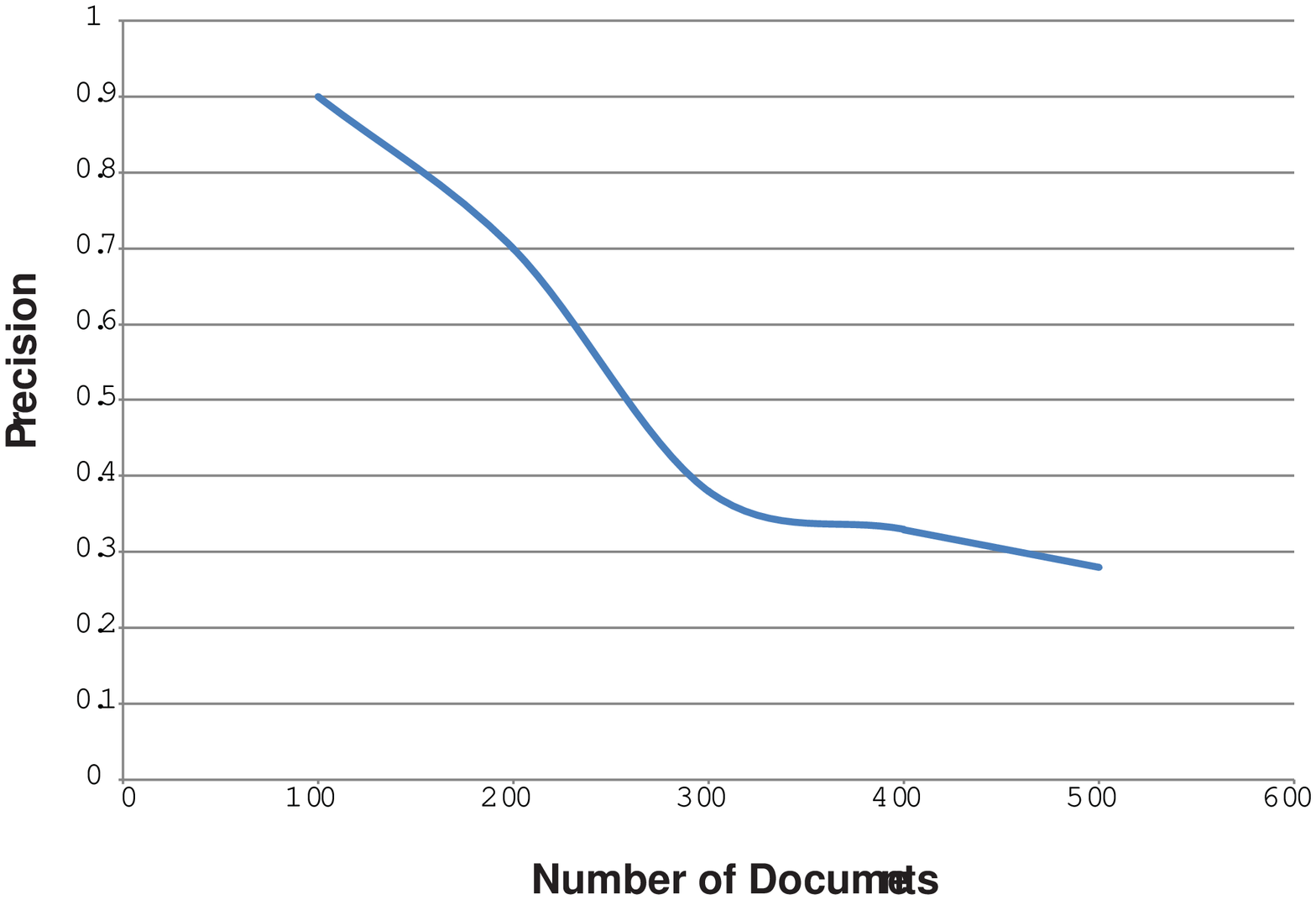, width=4in}
\caption{Precision Graph for Normal Queries}
\label{fig_q1}
\end{figure}

\begin{figure}[!h]
\centering
\epsfig{file=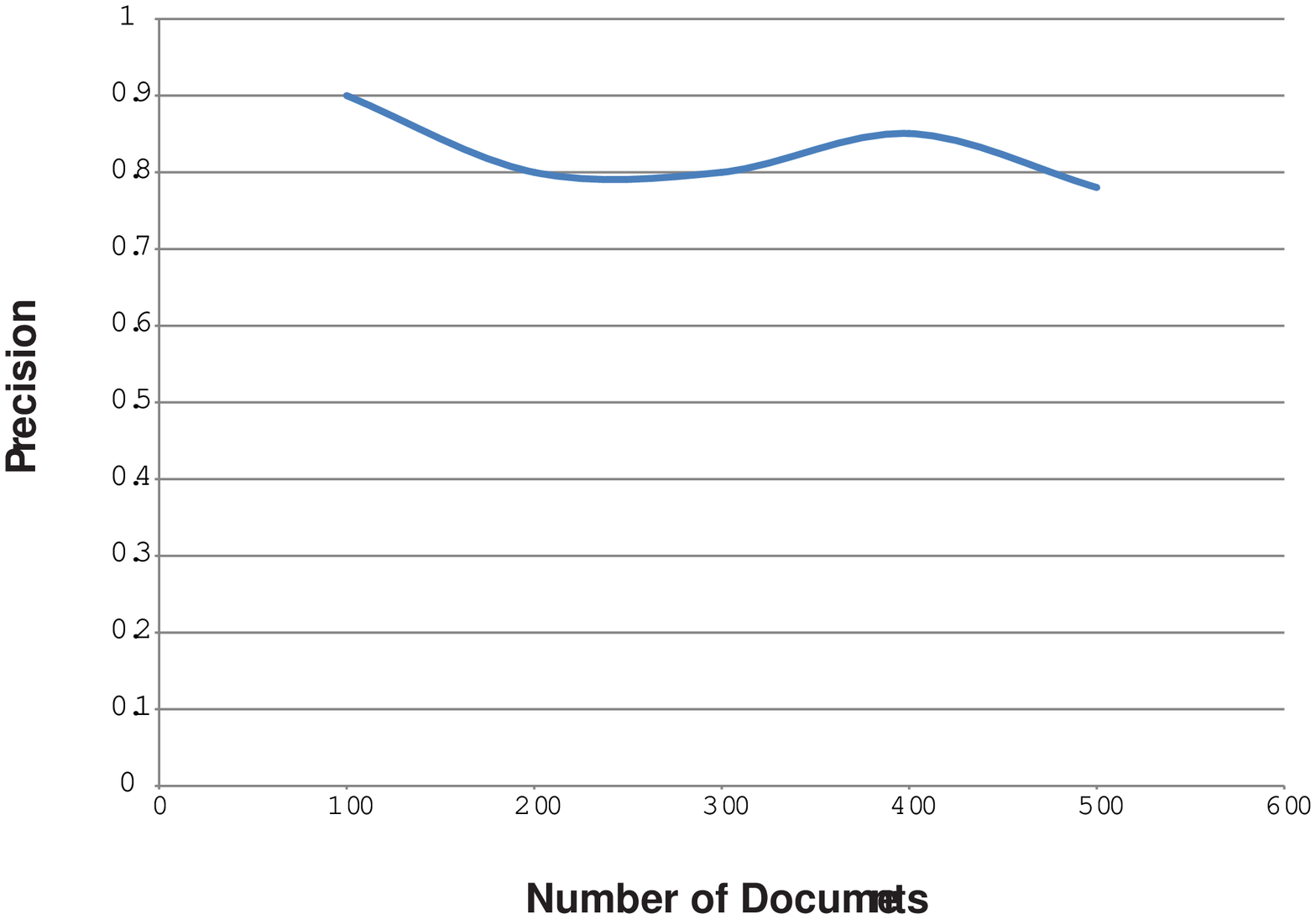, width=4in}
\caption{Precision Graph for Expanded Queries}
\vskip -2pt
\label{fig_q2}
\end{figure}

The precision for both normal query and expanded query were calculated for each increase in the total number of documents. The precision graphs are shown in Figure~\ref{fig_q1} and Figure~\ref{fig_q2}. 

The results indicate that, when the number of documents increase, the precision value for the expanded query decreases. This variation in the precision value is due to the fact that, at first, when the number of documents pertaining to each topic is smaller, the news articles matching the terms \textit{`kill'}, \textit{`out'} and \textit{`eliminate'} will be less as the news articles may not contain these terms at all or only topics like \textit{`Crimes'} or \textit{`Law Enforcement'} are present in the database.

Later, if some other topics like \textit{`Cricket'} or \textit{`Tennis'} or even \textit{`Politics'} are added, these topics also may contain the same keywords as in the expanded query but not in the same context. For instance, in the topic \textit{`Cricket'}, the keyword \textit{`out'} may occur, which has an entirely different meaning.  This makes them irrelevant news articles, which in turn causes the rapid fall in precision. There may be a slight increase in the precision if at some trail new news documents of topics like \textit{`Crime'} are added again.

In the case of normal query, the precision values did not show any drastic variations. This is because the evaluation occurs only by matching the word \textit{`kill'} and not the others, this intern gives the resultant news in more or less the same context. The word \textit{`kill'} would not always occur in different topics in different contexts. It will occur in different topics like \textit{`Politics'} or \textit{`Tennis'} only if it is a news about some political leader or a tennis player, actually, getting killed by someone or killing someone else. But sometimes, these retrieved news articles may become irrelevant if the user is looking for killings related to humans alone and not animals or birds.

From this, we can infer that, for getting very relevant news articles or for known-item search or navigational search, normal querying is the best way. But for getting a wide spectrum of news articles or to retrieve possibly linked news stories or if the users are not very sure of what they are looking for then, expanded querying can be used.

\subsection{Comparisons of Retrieval done using Original News Index, Pre-processed News Index and without using Lucene-based Index}

\begin{figure}[!h]
\centering
\epsfig{file=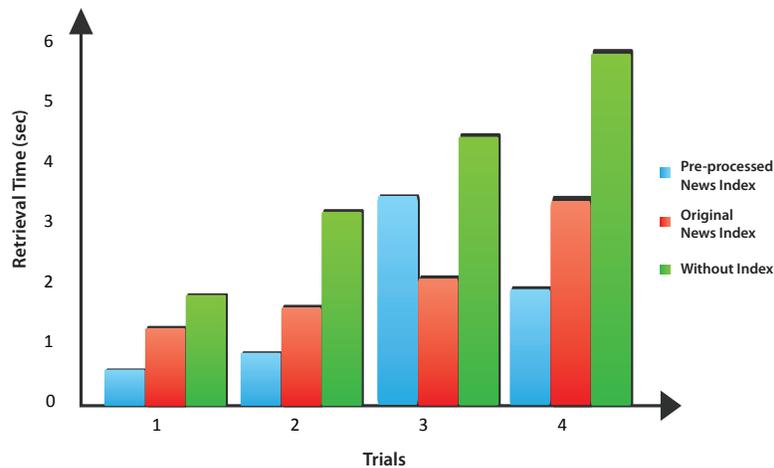, width=4in}
\caption{Retrieval Times for Various Methods of Retrieval}
\label{fig_indexchart}
\end{figure}

For this evaluation, four different trials are done. First trial was for exact query term matching and the second trial was for retrieval involving AND/OR/NOT evaluation. The third trial was for proximity search and retrieval and finally, the fourth trial was for retrieval involving wildcard character evaluation. The results are shown in the Figure~\ref{fig_indexchart}.

The results show that, for the first, second and fourth trials the retrieval time is more when retrieval is performed using the original index (index created without pre-processing the news articles). This is because the number of comparisons done are large, as it includes both stop words and non-stemmed words. 

But in the third trial, that is, when proximity search is performed using the original index, the retrieval time is lesser compared to the one performed using pre-processed index. This happens because the positional details of all the words in the news articles are stored in the original index. But it is lost when pre-processing is done before indexing the news articles. 

It is very clear that without using Lucene index, the retrieval takes very long and another advantage of Lucene-based index is its capability to be loaded into the database or filesystem or even the RAM of the system, which makes it flexible and efficient. Thus, retrieval done using Lucene-based index yields faster and efficient results for the input queries.

\section{Conclusion}

The next generation news retrieval system presented here, has met most of the pitfalls of today's news retrieval systems, such as, scalability, reliability and fault tolerance. The parallel news crawler used in this system is faster than the traditional crawlers, as it is designed using MapReduce programming model and is powered using Cloud technology. 

The parallel processing of news articles in MapReduce fashion ensures a hike in performance and also makes the news retrieval system more robust. The use of distributed database meets the huge storage space needed to store the evergrowing amount of news articles. The system is made fault tolerant, as replication of data is done after every write operation performed on the distributed database, automatically.

The Lucene-based indexing offers to reduce the retrieval time by half. The query expansion feature and the extractive summarization presented here, makes the retrieval more efficient. The 3D visualization of the news articles makes this news retrieval system more interactive and engaging. Features like the processing and retrieval of news in the form of image or video or audio in addition to the news articles in textual format and automatic backup tasks can be thought as future enhancements.


\end{document}